\documentclass[lettersize,journal]{IEEEtran}
\usepackage{amsmath,amsfonts}
\usepackage[noend]{algpseudocode}
\usepackage{array}
\usepackage{textcomp}
\usepackage{stfloats}
\usepackage{url}
\usepackage[hidelinks,colorlinks=true,linkcolor=blue,citecolor=blue,urlcolor=black]{hyperref}
\usepackage{acronym}
\usepackage{array}
\usepackage{tabularray}
\usepackage{verbatim}
\usepackage{subfigure}
\usepackage{graphicx}
\usepackage[dvipsnames]{xcolor}
\usepackage{cite}
\usepackage{multirow}
\usepackage{multicol}

\input{AcronymList}
\usepackage[linesnumbered,ruled,vlined]{algorithm2e}

\SetCommentSty{mycommfont}

\begin{document}

\title{A Novel Pilot Allocation Technique for Uplink OFDMA in ISAC Systems}

\author{Ahmet Sacid S\"{u}mer, Ebubekir Memi\c{s}o\u{g}lu and H\"{u}seyin Arslan,~\IEEEmembership{Fellow,~IEEE,}
\thanks{The authors are with the Department of Electrical and Electronics Engineering, Istanbul Medipol University, Istanbul, 34810, Turkey (e-mail: ahmet.sumer@std.medipol.edu.tr; ememisoglu@medipol.edu.tr; huseyinarslan@medipol.edu.tr). \\
}
}

\markboth{Journal of \LaTeX\ Class Files,~Vol.~14, No.~8, August~2021}%
{Shell \MakeLowercase{\textit{et al.}}: A Sample Article Using IEEEtran.cls for IEEE Journals}

\maketitle
\begin{abstract}
In integrated sensing and communication (ISAC) systems, pilot signals play a crucial role in enhancing sensing performance due to their strong autocorrelation properties and high transmission power. However, conventional interleaved pilots inherently constrain the maximum unambiguous range and reduce the accuracy of channel impulse response (CIR) estimation compared to continuous orthogonal frequency-division multiple access (OFDMA) signals. To address this challenge, we propose a novel overlapped block-pilot  structure for uplink OFDMA-based ISAC systems, called phase-shifted ISAC (PS-ISAC) pilot allocation. The proposed method leverages a cyclic prefix (CP)-based phase-shifted pilot design, enabling efficient multi-transmitter pilot separation at the receiver. Simulation results confirm that the proposed scheme enhances CIR separation, reduces computational complexity, and improves mean square error (MSE) performance under practical power constraints. Furthermore, we demonstrate that utilizing continuous pilot resources maximizes the unambiguous range.
\end{abstract}


\section{Introduction}
\IEEEPARstart{I}{ntegrated} sensing and communication (ISAC) has garnered significant attention from academia and industry due to its potential to enhance spectral, energy, and hardware efficiency by sharing spectrum and hardware resources for both sensing and communication functions \cite{liu2022integrated}. A key challenge in ISAC is designing \ac{MA} techniques that balance communication and sensing while minimizing mutual interference \cite{liu2024next,tusha2024interference}. In this context, \ac{NGMA} techniques are expected to play a crucial role in 6G wireless networks by optimizing resource allocation for dual functionality. From a resource allocation perspective, \ac{MA} techniques must balance the trade-off between sensing and communication \cite{liu2022evolution}. However, conventional \ac{MA} approaches, designed primarily for communication, may not be directly applicable to ISAC due to their inherent constraints in pilot allocation, interference management, and coexistence strategies.

In ISAC, pilot allocation plays a pivotal role in balancing the coexistence of sensing and communication signals. Traditionally, pilots have been employed in communication for channel estimation and positioning \cite{ozdemir2007channel}, while also serving as essential components in sensing applications \cite{5G_PRS}. Given this dual functionality, different wireless standards have adopted distinct pilot allocation strategies to optimize performance.
Recognizing this, Wi-Fi integrates sensing by transmitting pilot-only packets, utilizing the same channel access mechanisms as communication packets \cite{Wifi_sensing}. Meanwhile, in OFDMA-based systems, pilot allocation follows a structured approach, broadly classified into two categories: block and interleaved pilot allocation \cite{OFDMA_survey}.
A promising approach is \ac{UL} overlapped interleaved pilot allocation, where time-shifted pilot sequences are leveraged to separate the \acp{CIR} of multiple transmitters in the time domain \cite{Ribeiro2008UplinkCE}. Additionally, to mitigate excessive pilot overhead, overlapping pilots have been explored in prior studies \cite{overlapping_pilots_2005}, where Kalman estimators and predictive filtering techniques have been used to enhance the channel estimation process in \ac{UL} overlapped interleaved allocation. However, these methods do not account for the unique requirements of ISAC, such as a high maximum unambiguous range and velocity estimation.

\textbf{ISAC pilot designs:}
In \ac{OFDMA} systems, pilot signals are typically non-continuous and uniformly interleaved across the time and frequency domains. Beyond communication, these structured pilots have also been explored for sensing applications. Radar processing techniques leveraging pilots in the \ac{OFDM} waveform, along with dynamic pilot subcarrier allocation for range-adaptive radar sensing, have been investigated in the \ac{DL} \cite{Ozkaptan_1, Ozkaptan_2}. Additionally, the potential use of 5G \ac{PRS} for radar sensing in the \ac{DL} has been explored in \cite{5G_PRS}.
However, conventional pilot allocation schemes, originally designed for communication, suffer from reduced resolution and limited unambiguous range due to their interleaved spectrum occupation. To address this limitation, a coprime and periodic pilot design was proposed in \cite{Coprime_ISAC} as an interleaved pilot allocation strategy for the \ac{DL}. While this approach enhances the maximum unambiguous range, it introduces additional computational complexity. Furthermore, in the \ac{UL}, ISAC-oriented pilot allocation techniques have been investigated in the context of overlapped pilot and sensing sequences \cite{memisoglu2023csi, demir2023csi}. These methods primarily rely on interleaved pilot structures and iterative channel estimation, demonstrating effectiveness in sensing but also increasing computational complexity.

Despite the growing interest in ISAC pilot allocation, there remains a lack of research on block pilot structures in an overlapped manner for \ac{UL} \ac{OFDMA}-based ISAC systems. Current interleaved methods suffer from low pilot density, which degrades channel estimation accuracy and restricts the maximum unambiguous range. Therefore, there is a pressing need for a novel \ac{MA} strategy for \ac{UL} pilot allocation that overcomes these challenges without introducing excessive complexity. Such a solution must maximize the unambiguous range limit, enhance sensing accuracy, minimize computational overhead, and ensure robust multi-transmitter \ac{CIR} separation in overlapped pilot allocation schemes for \ac{OFDMA}-based ISAC systems.

\textbf{Contributions of this work:}
In this work, we propose a novel overlapped block-pilot structure for \ac{UL} \ac{OFDMA}-based ISAC systems, introducing a \ac{CP}-based phase-shifted pilot allocation to enable efficient multi-transmitter pilot separation at the receiver. The proposed \ac{PS-ISAC} method ensures continuous pilot transmission, effectively enhancing the maximum unambiguous range while optimizing spectral utilization. Performance evaluation confirms that \ac{PS-ISAC} outperforms conventional interleaved pilot allocation—widely used in standardized \ac{OFDMA} systems \cite{3gpp2025_38211}—by achieving lower \ac{MSE} under power spectral constraints with reduced computational complexity. These results establish the overlapped block-pilot structure as a robust and efficient alternative to standardized orthogonal interleaved schemes, offering enhanced estimation accuracy and resource utilization for ISAC systems.

\section{System Model}
\begin{figure*}[t]
    \centering
    \includegraphics[width=1.0\linewidth]{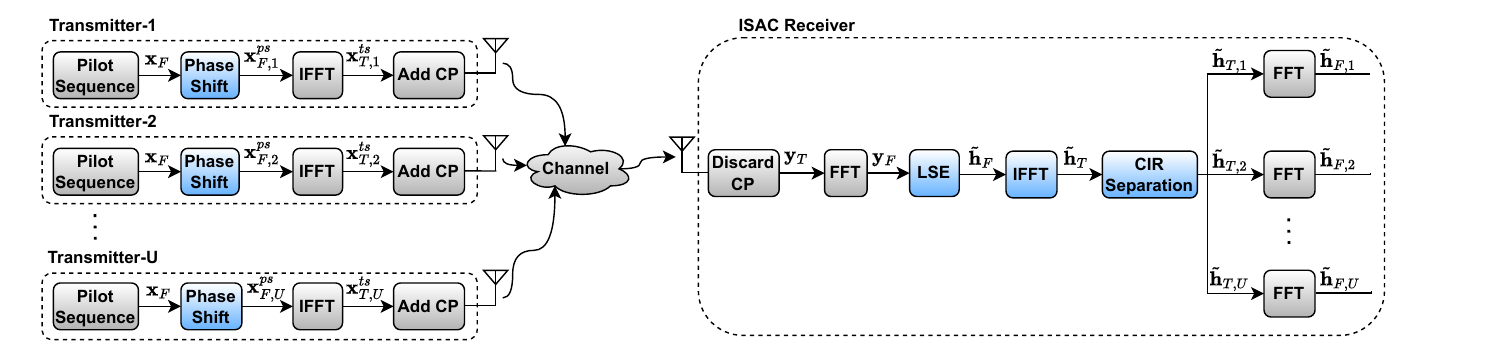}
    \caption{The system model of the proposed uplink ISAC multiple-access scheme.}
    \label{fig:system_model}
\end{figure*}
The proposed system model, illustrated in Fig.~\ref{fig:system_model}\footnote{The blocks differing from \ac{CI-ISAC} in the proposed \ac{PS-ISAC} method are highlighted in blue.}, consists of $U$ single-antenna transmitters, each functioning as either a communication or sensing transmitter in an \ac{UL} \ac{OFDMA} transmission. These transmitters simultaneously transmit their signals over a channel to a single-antenna ISAC receiver, sharing the same time and frequency resources to maximize spectrum efficiency. Each OFDM symbol is structured based on the transmitter type: communication transmitters use pilot carriers, while sensing transmitters use dedicated sequences\footnote{Since sensing sequences incorporate known pilot symbols for channel estimation, both sensing and communication transmitters can be treated similarly at the receiver. Thus, they are collectively referred to as transmitters, and both pilot and sensing sequences are simply termed pilots \cite{Ozkaptan_1}.}. 
During transmission, pilots $\mathbf{x}_F$\footnote{Boldface lowercase symbols represent vectors.} are inserted by the transmitters, followed by a phase shift applied to the frequency-domain signal based on the information provided by the receiver. 
Subsequently, the time-domain \ac{OFDM} symbol is generated as

\begin{equation}
    \begin{aligned}
    \label{eq:txsignal_time}
    x_{T,u}^{ts}(n) \!& =\! \frac{1}{\sqrt{N}} \sum_{k=0}^{N-1} x_{F,u}^{ps}(k) e^{j 2 \pi k n / N}, \quad \!\!\! n \!= \!0,1, \dots, N\!-\!1,
    \end{aligned}
\end{equation}
where $N$ represents the \ac{FFT} size, and $x_{F,u}^{ps}(k)$ denotes the phase-shifted pilot in the frequency domain. Then, a \ac{CP} of size $N_{\text{CP}}$ is appended. The signal is then processed through digital-to-analog conversion and transmitted over the channel by each transmitter separately. 
While the same $N_{\text{CP}}$ is used for both communication and sensing symbols, different values may be assigned, provided  the maximum excess delay of the channel is less than $N_{\text{CP}}$.

At the ISAC receiver, the received signal comprises the summation of all transmitted signals. The reception process begins with analog-to-digital conversion, followed by \ac{CP} removal, yielding the time-domain signal $\mathbf{y}_T$. Subsequently, the received frequency-domain signal $\mathbf{y}_F$ can be expressed as
\begin{equation}
\label{eq:rxsignal_freq}
    y_F(k) = \sum_{u=1}^{U} h_{F,u}(k) x_{F,u}^{ps}(k) + w_u(k), \quad \!\!\! k \!= \!0,1,\dots, N\!-\!1,
\end{equation}
where \( h_{F,u}(k) \) represents the Rayleigh fading channel samples, modeled as \( \mathcal{CN}(0,1) \), and \( w_u(k) \) denotes an AWGN sample, modeled as \( \mathcal{CN}(0,\sigma^2) \).
After receiving the signal, channel estimation is performed, and the estimated \ac{CFR} is subsequently transformed into the time domain via an \ac{IFFT} operation for channel separation and noise cancellation \cite{ozdemir2007channel}.
The \ac{CIR} of each transmitter is individually extracted in the time domain, followed by an \ac{FFT} operation to recover the \ac{CFR} for each transmitter.  
For parallel transmission, all communication and sensing transmitters are assumed to be perfectly synchronized with the receiver in both time and frequency domains.

\begin{figure}
  \centering
     \subfigure[CIRs obtained using the PS-ISAC method.]
     {\includegraphics[width=0.3\textwidth]{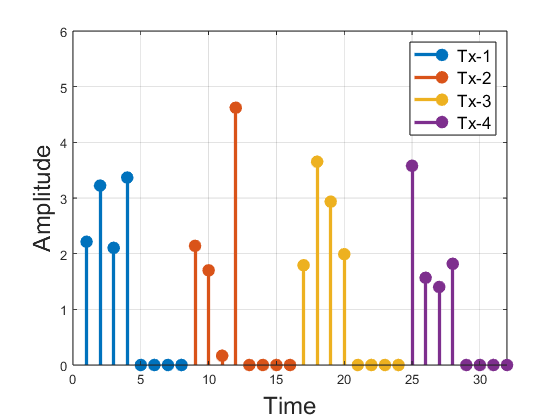}} 
     \subfigure[CIRs obtained using the CI-ISAC method.]{\includegraphics[width=0.5\textwidth]{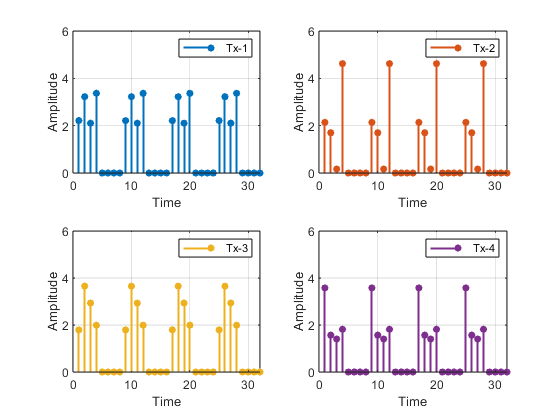}}  
  \caption{CIRs of transmitters at the receiver for different pilot design methods.}   
      \label{fig:CIRs}
\end{figure}

\begin{table*}[t]
\centering
\caption{Computational Complexity of CI-ISAC and PS-ISAC.} 
\label{table:complexity}
\begin{tabular}{l|ll|ll|}
\cline{2-5}
                                     & \multicolumn{2}{l|}{Transmitters}                                    & \multicolumn{2}{l|}{Receiver}                                                       \\ \cline{2-5} 
                                     & \multicolumn{1}{l|}{CI-ISAC}                & PS-ISAC               & \multicolumn{1}{l|}{CI-ISAC}                        & PS-ISAC                       \\ \hline
\multicolumn{1}{|l|}{Addition}       & \multicolumn{1}{l|}{$ \!\!\! U(3N\log_{2}N- 3N+4) \!\!\!$} & $ \!\!\!U(3N\log_{2}N- 3N+4) \!\!\!$ & \multicolumn{1}{l|}{$ \!\!\!(2U+1)(3N\log_{2}N- 3N+4) \!\!\!$}    & $ \!\!\!(U+2)(3N\log_{2}N- 3N+4) \!\!\!$   \\ \hline
\multicolumn{1}{|l|}{Multiplication} & \multicolumn{1}{l|}{$\!\!\!U(N\log_{2}N -3N+4)\!\!\!$}  & $\!\!\!U(N\log_{2}N -N+4)\!\!\!$  & \multicolumn{1}{l|}{$\!\!\!(2U+1)(N\log_{2}N -3N+4)+ 2N\!\!\!$} & $\!\!\!(U+2)(N\log_{2}N -3N+4)+ 2N\!\!\!$ \\ \hline
\end{tabular}
\end{table*}

\section{Proposed Method}

In \ac{CI-ISAC} for \ac{OFDMA} systems, pilots are utilized for both channel estimation and sensing \cite{Ozkaptan_2}, where the estimated channel is subsequently employed for either data demodulation or sensing tasks. However, in the proposed \ac{PS-ISAC}, signals are fully overlapped in both the time and frequency domains, making conventional channel estimation infeasible due to their inseparability. To address this limitation, phase-shifted pilot sequences are applied at the transmitters, allowing \ac{CIR} separation at the receiver after the \ac{IFFT} operation, where each transmitter's \ac{CIR} appears at a distinct $N_\text{CP}$ interval.

Prior to transmission, the receiver assigns each transmitter a phase shift based on $N_\text{CP}$, which is then applied to the pilot sequences in the frequency domain as  
\begin{equation}
\begin{aligned}
    \label{eq:phase_shifted}
    x_{F,u}^{ps}(k) & = \psi_u(k) x_{F}(k), 
\end{aligned}
\end{equation}
where the phase shift introduces a corresponding time shift in the time domain after the \ac{IFFT} operation. As illustrated in Fig.~\ref{fig:system_model}, each transmitter $u$ applies the phase shift  
\begin{equation}
\begin{aligned}
\label{eq:phase_shift_term}
 \psi_u(k) &= e^{-j 2 \pi k n_u /N}, \quad  k = 0,1,\dots, N-1,
    \end{aligned}
\end{equation}
where the phase shift amount is defined as $n_u = (u-1) N_\text{CP}$.  
Since $N_\text{CP}$ is known at both the transmitter and receiver, no additional signaling overhead is required for timing synchronization. The only requirement is to transmit the assigned phase shift to each transmitter, similar to the interleave subcarrier offset in conventional interleaved systems \cite{3gpp2025_38211}.

At the receiver, \ac{CFR} estimation is performed using the known pilots $\mathbf{x}_{F}$. The received frequency-domain signal from \eqref{eq:rxsignal_freq} allows \ac{CFR} estimation via the \ac{LS} estimator \cite{ozdemir2007channel}. Unlike \ac{CI-ISAC}, where \ac{LS} is performed for each transmitter, in \ac{PS-ISAC}, a single LS estimation is applied to all transmitters simultaneously. The estimated \ac{CFR} is given by  
\begin{equation}
\begin{aligned}
    \label{eq:FDE}
    \Tilde{h}_F(k) & = \frac{y_F(k)}{x_{F}(k)}, \quad k=1,2,\dots,N.
\end{aligned}
\end{equation}
To enhance estimation accuracy, the estimated \ac{CFR} is converted to \ac{CIR}, facilitating multi-transmitter \ac{CIR} separation as
\begin{equation}
\begin{aligned}
\label{eq:estimated_time_channel}
        \Tilde{h}_T(n) = \frac{1}{\sqrt{N}} \sum_{k=0}^{N-1} \Tilde{h}_F(k) e^{j 2 \pi k n / N}.
\end{aligned}
\end{equation}
Then, transmitter-specific \ac{CFR}s are extracted as
\begin{equation}
\begin{aligned}
    \label{eq:H_Tilde_FFT}
    \Tilde{h}_{F,u}(k) & = \sqrt{N} \sum_{n=(u-1) N_{\text{CP}}}^{u N_{\text{CP}}-1} \Tilde{h}_T(n) e^{- j 2 \pi n k / N},
\end{aligned}
\end{equation}
where the index selection in the \ac{FFT} summation defines the \ac{CIR} separation block, ensuring accurate differentiation of transmitter signals in the time domain, as shown in Fig.~\ref{fig:system_model}.  

Fig.~\ref{fig:CIRs} illustrates the time-domain separation of transmitters' \acp{CIR} at the receiver, highlighting the distinction between the proposed \ac{PS-ISAC} and \ac{CI-ISAC}.  
The parameters used are $N=32$, $U=4$, $N_\text{CP}=8$, a 4-tap frequency-selective Rayleigh fading channel, and \ac{PR} $= 1/U$ with pseudo-random pilot sequences. In \ac{PS-ISAC}, leveraging $N_\text{CP}$ segments the received signal into intervals, enabling \ac{CIR} separation in the time domain. 
Due to the overlapped block-pilot structure, periodicity does not appear in \ac{PS-ISAC}. Conversely, \ac{CI-ISAC} follows a fixed periodic structure, where interleaved pilots in the frequency domain introduce periodicity in time. This key difference allows \ac{PS-ISAC} to separate \acp{CIR} in the time domain, whereas in \ac{CI-ISAC}, \ac{CFR} separation is achieved using transmitter-specific interleaved pilots. 

\begin{figure}[t]
    \centering
    \includegraphics[width=1\linewidth]{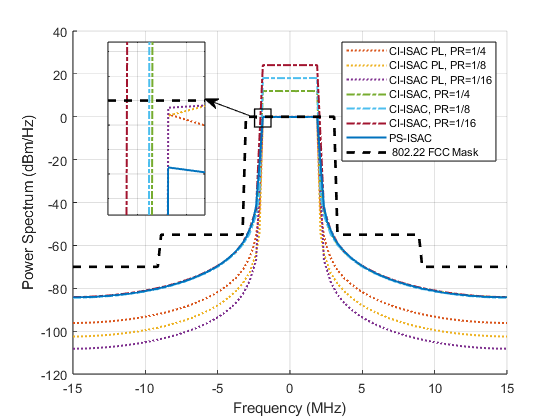}
    \caption{Power spectra of CI-ISAC and PS-ISAC transmitters.}
    \label{fig:PSOFDM}
\end{figure}

\subsection{Computational Complexity Analysis}
The computational complexity analysis considers real additions (add.)/subtractions, multiplications (mult.)/divisions performed at the transmitter and receiver over an \ac{OFDM} symbol duration. The \ac{FFT} operation requires $(3N\log_{2}N - 3N + 4)$ real additions and $(N\log_{2}N - 3N + 4)$ real multiplications \cite{sorensen1986computing}. At the transmitters, \ac{IFFT} complexity remains the same for both \ac{CI-ISAC} and \ac{PS-ISAC}, but \ac{PS-ISAC} introduces additional overhead due to the phase shift operation, requiring $2N$ real multiplications per transmitter.  
At the receiver, \ac{FFT} complexities for both \ac{CI-ISAC} and \ac{PS-ISAC} are identical when computing $\mathbf{y}_F$ and $\mathbf{\Tilde{h}}_{F,u}$ in \eqref{eq:H_Tilde_FFT}. Computing $\mathbf{y}_F$ requires a single \ac{FFT} for all transmitters, whereas obtaining $\mathbf{\Tilde{h}}_{F,u}$ requires an \ac{FFT} for each transmitter. Both methods require $2N$ real multiplications for $\mathbf{\Tilde{h}}_{F}$ estimation in \eqref{eq:FDE}.  
The key difference arises in \ac{IFFT} operations at the receiver. In \ac{CI-ISAC}, computing $\Tilde{\mathbf{h}}_{T,u}$ for each transmitter requires a separate \ac{IFFT}, whereas in \ac{PS-ISAC}, a single \ac{IFFT} suffices, making it independent of the number of transmitters. A computational complexity comparison is provided in Table~\ref{table:complexity}.

\begin{figure}[t]
    \centering
    \includegraphics[width=1\linewidth]{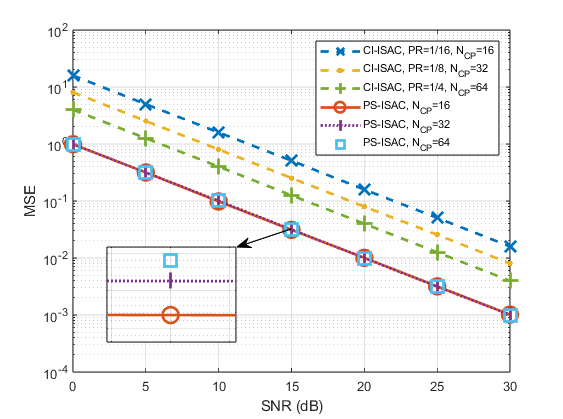}
    \caption{MSE performance of CI-ISAC and PS-ISAC under power constraints.}
    \label{fig:MSEwPL}
\end{figure}

\begin{figure}[t]
    \centering
    \includegraphics[width=1\linewidth]{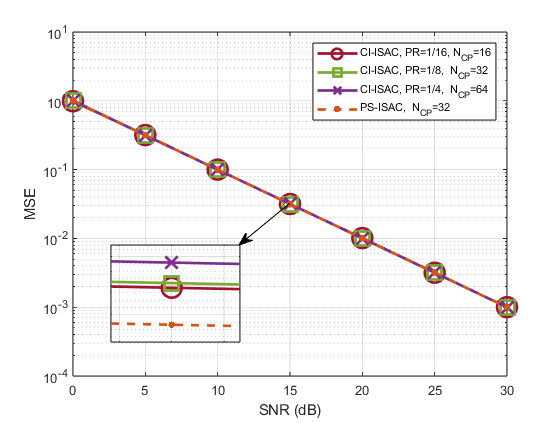}
    \caption{MSE performance of CI-ISAC and PS-ISAC without power constraints.}
    \label{fig:MSEwoPL}
\end{figure}

\subsection{Maximum Unambiguous Range}

    The maximum unambiguous range is given by \cite{Coprime_ISAC}  
\begin{equation}
    R_{\max} = \frac{N_{p} c}{2 N \Delta f}, 
\end{equation}
where \( N_{p} \) is the number of pilot-carrying subcarriers, \( \Delta f \) is the subcarrier spacing, and \( c \) denotes the speed of light. In the interleaved structure, \( N_{p} \) is given by \( N_{p} = N \times \text{PR} \), while in the proposed overlapped block-pilot method, it simplifies to \( N_{p} = N \) since PR is set to 1. This confirms that the proposed method fully utilizes all available subcarriers for pilot allocation, achieving the theoretical limit for the maximum unambiguous range.

\section{Simulation Results}
In this section, Monte Carlo simulations evaluate the \ac{MSE} performance of \ac{PS-ISAC}, alongside computational complexity and maximum unambiguous range analysis for a varying number of transmitters.  
The simulations use pseudo-random pilot sequences with \( N = 256 \) and \( N_{\text{CP}} = N \times \text{PR} \). Channels are modeled as \( (N_{\text{CP}} - 1) \)-tap frequency-selective Rayleigh fading, with \( U = 1/\text{PR} \).  

Fig.~\ref{fig:PSOFDM} examines the impact of \ac{PC}, comparing the PS-ISAC (PR = 1) with the CI-ISAC (PR = 1/4, 1/8, 1/16) under \ac{PC} or without \ac{PC}. 
In \ac{PC} scenarios, each pilot subcarrier is normalized to unit power, while in non-\ac{PC} cases, pilot subcarriers are scaled by $\sqrt{\text{1/PR}}$ to maintain total power consistency with PR=1. However, as shown in Fig.~\ref{fig:PSOFDM}, the non-\ac{PC} approach exceeds the 802.22 FCC spectrum mask \cite{tom2013mask}, while \ac{PS-ISAC} and \ac{CI-ISAC} under \ac{PC} remain within the mask due to uniform power distribution across pilot carriers.

Figs.~\ref{fig:MSEwPL} and \ref{fig:MSEwoPL} compare \ac{MSE} performances of PS-ISAC and CI-ISAC in \ac{PC} and non-\ac{PC} scenarios for different PR and $N_{\text{CP}}$ values. The \ac{MSE} is computed as $\frac{1}{U} \sum_{u=1}^U \mathbb{E}[(\mathbf{h}_{F,u}-\Tilde{\mathbf{h}}_{F, u})^2]$. \ac{PS-ISAC} achieves superior or comparable \ac{MSE} performance to \ac{CI-ISAC}, depending on the application of power constraints.   
Fig.~\ref{fig:MSEwPL} examines \ac{PS-ISAC} under regulatory \ac{PC}, showing consistent superiority over \ac{CI-ISAC} across all \acp{PR}. Since \ac{PS-ISAC} allows all transmitters to fully utilize subcarriers (\ac{PR}=1), \ac{MSE} remains independent of the number of transmitters. In contrast, \ac{CI-ISAC} experiences degraded \ac{MSE} as the number of transmitters increases relative to PR. The only trade-off in PS-ISAC is its higher pilot power usage per transmitter compared to CI-ISAC.
Fig.~\ref{fig:MSEwoPL} illustrates the \ac{MSE} performance without \ac{PC}, showing that \ac{PS-ISAC} achieves the same \ac{MSE} as \ac{CI-ISAC}. Since the \ac{MSE} remains unchanged for any \( N_{\text{CP}} \) value due to the overlapped block-pilot allocation, only \( N_{\text{CP}} = 32 \) is presented for \ac{PS-ISAC}.  

\begin{table}[]
\centering
\caption{Computational complexity comparison between CI-ISAC and PS-ISAC.}
\label{table:complexity_calc}
\begin{tabular}{cc|cc|cc|}
\cline{3-6}
\multicolumn{1}{l}{}                           & \multicolumn{1}{l|}{} & \multicolumn{2}{c|}{Transmitters}   & \multicolumn{2}{c|}{Receiver}       \\ \hline
\multicolumn{1}{|c|}{Method}                   & U                     & \multicolumn{1}{c|}{Add.}  & Mult. & \multicolumn{1}{c|}{Add.}   & Mult. \\ \hline
\multicolumn{1}{|c|} {\multirow{3}{*}{CI-ISAC}} & 4                     & \multicolumn{1}{c|}{21520} & 5136  & \multicolumn{1}{c|}{48420}  & 12068 \\ \cline{2-6} 
\multicolumn{1}{|c|}{}                         & 8                     & \multicolumn{1}{c|}{43040} & 10272 & \multicolumn{1}{c|}{91460}  & 22340 \\ \cline{2-6} 
\multicolumn{1}{|c|}{}                         & 16                    & \multicolumn{1}{c|}{86080} & 20544 & \multicolumn{1}{c|}{177540} & 42884 \\ \hline
\multicolumn{1}{|c|}{\multirow{3}{*}{PS-ISAC}} & 4                     & \multicolumn{1}{c|}{21520} & 7184  & \multicolumn{1}{c|}{32280}  & 8216  \\ \cline{2-6} 
\multicolumn{1}{|c|}{}                         & 8                     & \multicolumn{1}{c|}{43040} & 14368 & \multicolumn{1}{c|}{53800}  & 13352 \\ \cline{2-6} 
\multicolumn{1}{|c|}{}                         & 16                    & \multicolumn{1}{c|}{86080} & 28736 & \multicolumn{1}{c|}{96840}  & 23624 \\ \hline
\end{tabular}
\end{table}

The computational complexity calculations for a varying number of transmitters are summarized in Table~\ref{table:complexity_calc}. Results show that PS-ISAC consistently has lower computational complexity than CI-ISAC, with the difference becoming more significant as the number of transmitters increases. This is due to \ac{PS-ISAC} requiring only a single \ac{IFFT} at the receiver for \ac{CIR} estimation. While transmitter complexity is slightly higher due to the phase shift operation, it remains negligible compared to the computational cost of \ac{IFFT}, especially for large values of $N$.  

\begin{table}[]
\centering
\caption{Maximum unambiguous range comparison between CI-ISAC and PS-ISAC.} 
\label{table:MUR_calc}
\begin{tabular}{|c|c|c|}
\hline
Method                   & U      & $R_{\max}$(m) \\ \hline
\multirow{3}{*}{CI-ISAC} & 4      & 2498   \\ \cline{2-3} 
                         & 8      & 1249   \\ \cline{2-3} 
                         & 16     & 624    \\ \hline
PS-ISAC                  & 4/8/16 & 9993   \\ \hline
\end{tabular}
\end{table}

Table~\ref{table:MUR_calc} highlights the maximum unambiguous range advantage of \ac{PS-ISAC}. When PR=1, the proposed method achieves the highest possible $R_{\max}$, whereas in \ac{CI-ISAC}, $R_{\max}$ decreases as the number of transmitters increases.   
Overall, \ac{PS-ISAC} not only maximizes the maximum unambiguous range with significantly lower computational complexity but also achieves superior \ac{MSE} performance under power constraints while maintaining equivalent \ac{MSE} when power constraints are not applied. These results underscore the efficiency of \ac{PS-ISAC} in resource-constrained environments.

\section{Conclusion}
This paper proposes a novel overlapped block-pilot allocation method for uplink OFDMA-based ISAC systems, leveraging a CP-based phase-shifted pilot design to enable the separation of the CIRs from multiple transmitters at the receiver. Theoretical analysis and simulations confirm that PS-ISAC substantially reduces computational complexity by eliminating the need for transmitter-specific IFFT operations at the receiver, compared to the CI-ISAC method. Additionally, it maintains the highest achievable maximum unambiguous range and enhances the MSE performance under power constraints. These advantages make PS-ISAC as a scalable and efficient pilot design solution for next-generation ISAC networks.



\vfill

\end{document}